Cryogen-Free dissolution Dynamic Nuclear Polarization polarizer operating at 3.35 T, 6.70 T and 10.1 T.

**Authors:**

Jan Henrik Ardenkjær-Larsen[1,2], Sean Bowen[1A], Jan Raagaard Petersen[1], Oleksandr Rybalko[1], Mads Sloth Vinding[3B], Marcus Ullisch[3] and Niels Chr. Nielsen[3]

**Affiliations:**

[1]Department of Electrical Engineering, Center for Hyperpolarization in Magnetic Resonance, Technical University of Denmark, Kgs. Lyngby, Denmark

[2]GE Healthcare, Brøndby, Denmark.

[3]Interdisciplinary Nanoscience Center (iNANO) and Department of Chemistry, Aarhus University, Aarhus, Denmark

**Corresponding Author and Reprint info:**

Jan Henrik Ardenkjær-Larsen, DTU Electrical Engineering, Ørsteds Plads, Building 349, room 126, 2800 Kgs. Lyngby, Denmark. Tel (0045) 4525 3918. E-mail: jhar@elektro.dtu.dk

**Grant support:**

This work was supported by the Danish Council for Independent Research (1323-00331B/00331A) and the Danish National Research Foundation (grant number DNRF124).

The authors confirm that this article content has no conflict of interest.

Word count: 5110

**Present address:**

[A] Merck & Co. Inc., West Point, PA, USA

[B] Center of Functionally Integrative Neuroscience, Aarhus University, Aarhus, Denmark

**Abstract**

Purpose: A novel dissolution dynamic nuclear polarization (dDNP) polarizer platform is presented. The polarizer meets a number of key requirements for *in vitro*, pre-clinical and clinical applications.

Method: It uses no liquid cryogens, operates in continuous mode, accommodates a wide range of sample sizes up to and including those required for human studies, and is fully automated.

Results: It offers a wide operational window both in terms of magnetic field, up to 10.1 T, and temperature, from room temperature down to 1.3 K. The polarizer delivers a $^{13}$C liquid state polarization for [1-$^{13}$C]pyruvate of 70%. The build-up time constant in the solid state is approx. 1200 s (20 min), allowing a sample throughput of at least one sample per hour including sample loading and dissolution.

Conclusion: We confirm the previously reported strong field dependence in the range 3.35 to 6.7 T, but see no further increase in polarization when increasing the magnetic field strength to 10.1 T for [1-$^{13}$C]pyruvate and trityl. Using a custom dry magnet, cold head and recondensing, closed-cycle cooling system, combined with a modular DNP probe, automation and fluid handling systems; we have designed a unique dDNP system with unrivalled flexibility and performance.

**Introduction**

Despite the widespread utility of NMR and MRI, emerging applications are often limited by low sensitivity. To improve the sensitivity, it is possible to either increase the available signal or decrease noise levels, or both. Improving signal by stronger magnetic field is challenging, since superconducting magnet technology permits only relatively modest increases in magnetic field, which come at relatively high cost [1,2]. On the other hand, noise reduction is possible, chiefly driven by the development of cryogenically cooled probes [3]. However, this leads to a relatively modest factor of noise reduction of about a factor of approx. four. It is clear that an alternative scheme is necessary to make dramatic change to the signal-to-noise ratio. Using hyperpolarization, the sensitivity can be augmented by several orders of magnitude by boosting the polarization of the nuclear spins close to unity [4,5].

One of the leading means of hyperpolarization is dynamic nuclear polarization (DNP). DNP itself was predicted by Overhauser in 1953 and subsequently experimentally confirmed using paramagnetic metals [6–8]. The premise of DNP is the use of microwave (MW) irradiation to transfer polarization from electron spins with larger polarization to nuclear spins. While early work used the conduction electrons of metals, modern work uses exogenous stable free radicals, such as e.g. nitroxides or trityls. There are four different DNP mechanisms that drive the polarization transfer, namely the Overhauser effect (OE), cross effect (CE), solid effect (SE), and thermal mixing (TM) [9]. These mechanisms have different dependency on magnetic field strength and temperature. It is well established that for most samples at low temperature (<4.2 K), TM is the dominant mechanism. With increasing magnetic field strength, the ESR line width typically increases due to g-anisotropy, and the resonance becomes more inhomogeneously broadened. It is therefore well established that at higher magnetic field strength, modulation of the MW frequency [10] to saturate more electron spin packets or increasing the radical concentration to make the line more homogeneous [11], improves the DNP efficiency. The other DNP mechanisms should not be relevant in this study, but will also depend on the magnetic field strength. Nuclear and electron spin relaxation times will also depend on both magnetic field strength and temperature, and both play an important role in the efficiency of DNP. Griffin et al. pioneered the use of DNP with solid-state NMR at MIT, where it has been applied to a variety of systems [12–14]. One limitation of working in the solid-state, at low temperature is the lack of information on dynamics and chemical reactions. Therefore, it is desirable to get the best of both worlds, by performing DNP at low temperature, while observing in the liquid-state at ambient temperature. This method known as dissolution DNP (dDNP) involves the rapid dissolution of the frozen sample into hot solvent [15].

A key factor, which impairs the widespread implementation of dDNP, is the need for copious quantities of liquid helium for maintaining the sample at temperatures below 2 K. Helium is a costly, non-renewable resource, and dDNP systems typically consume more than 2 L of liquid helium per sample. While this can be offset by the use of helium recovery systems, these are also costly, and 100% recovery is quite difficult to achieve. Equally important is the inconvenience of frequent handling of cryogens in a hospital environment, or even in a typical NMR laboratory.

While a system for clinical use has been designed with zero cryogen consumption [16], the goals for a research system differ. Our aims were:

- Zero cryogen consumption with ability to run in continuous mode
- Fully automated operation for routine use
- Dual channel solid-state NMR probe with broadband capability
- Base temperature of less than 1.5 K under DNP conditions and operational at any temperature up to room temperature

- Operational at multiple field strengths up to 10.1 T corresponding to a MW frequency of 282 GHz
- Any sample size up to 2 g corresponding to a human dose of [1-$^{13}$C]pyruvic acid

The basic principles of the polarizer were first presented at the 5$^{th}$ International DNP Symposium, Egmond aan Zee, Netherlands, Sep 2015 [17], which inspired others to adopt a similar design [18]. More recently, we have presented preliminary results at ENC [19] and EUROMAR [20].

In this work, we present a detailed description of the HYPERMAG polarizer, and use the unique capability of the polarizer to study the DNP enhancement as a function of magnetic field strength for the most important imaging agent for hyperpolarized metabolic MR, [1-$^{13}$C]pyruvate [21,22]. Hyperpolarized pyruvate is now in man and first clinical studies have been published [23–26].

**Methods**

A photo of the HYPERMAG polarizer is shown in Fig 1. Central elements are labelled and referenced in the following sections.

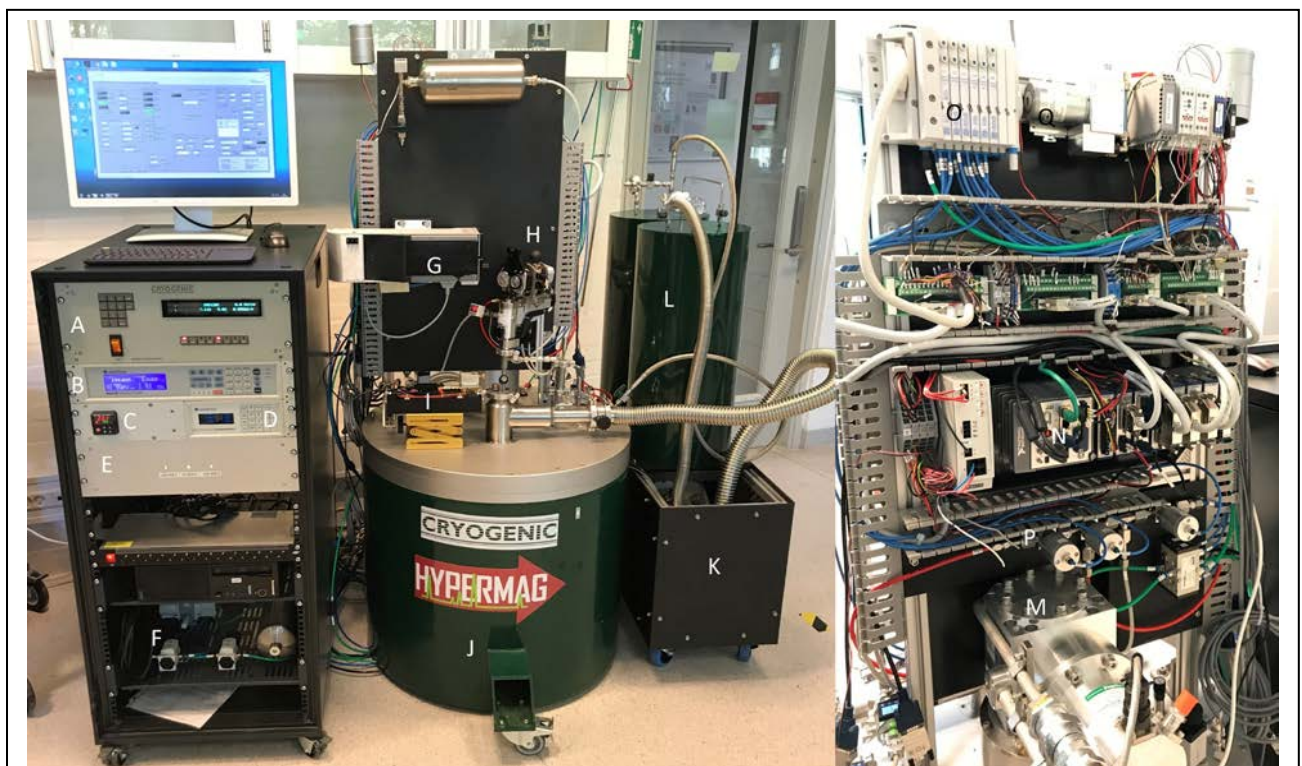

**Figure 1**: Photographs of HYPERMAG polarizer: A: magnet power supply; B: LS350 temperature controller; C: Watlow temperature controller for solvent heating; D: LS218 temperature logger; E: power supply; F: pneumatics; G: heater pressure module for fluid path syringe; H: Insertion module for fluid path, air lock and gate valve on top of DNP probe; I: MW source; J: dry magnet and cryostat; K: dry pump; L: buffer tanks; M: cold head; N: National Instruments cRIO controller; O: Pneumatics valve block for vacuum and pressure; P: high pressure valves for syringe drive; Q: vacuum pump for air lock.

*Cryostat*

Fig 2 shows a schematic of the cryostat (Fig 1J, Cryogenic Ltd, UK). The cooling power of the system is provided by a 1 W pulse tube cooler (Fig 1M: RP-082B2, SHI Cryogenics, Japan) and a F-70H compressor (SHI Cryogenics, Japan). A temperature monitor (Fig 1D: Model 218, Lakeshore, USA) monitors the magnet, internal components and cold head temperatures. The Variable Temperature Insert (VTI) is 30 mm diameter and 400 mm long. An oil-free, multi-stage roots blower with a maximum pumping speed of 37 m$^3$/h (Fig 1K: ACP40, Pfeiffer Vacuum, Germany) pumps the VTI. The exhaust of the pump is connected to a buffer volume of 100 L (Fig 1O) that is further connected to an inlet at the cryostat that leads to a liquid helium pot through condensation stages at the cold head. The flow of liquid helium to the VTI is regulated manually by a needle valve. The VTI has a Cernox temperature sensor (CX1030, Lakeshore, USA) and 50 W heater connected to a temperature controller (Fig 1B: Model 350, Lakeshore, USA). The VTI can operate continuously from base temperature to 300 K with the magnet at full field. The VTI pumping line and VTI condenser inlet are fitted with digital pressure transducers (910 DualTrans, MKS, Denmark). If the inlet pressure exceeds 700 mbar, electronic reduction of the pumping speed of the dry pump protects the magnet from an excessive heat load and potential quenching.

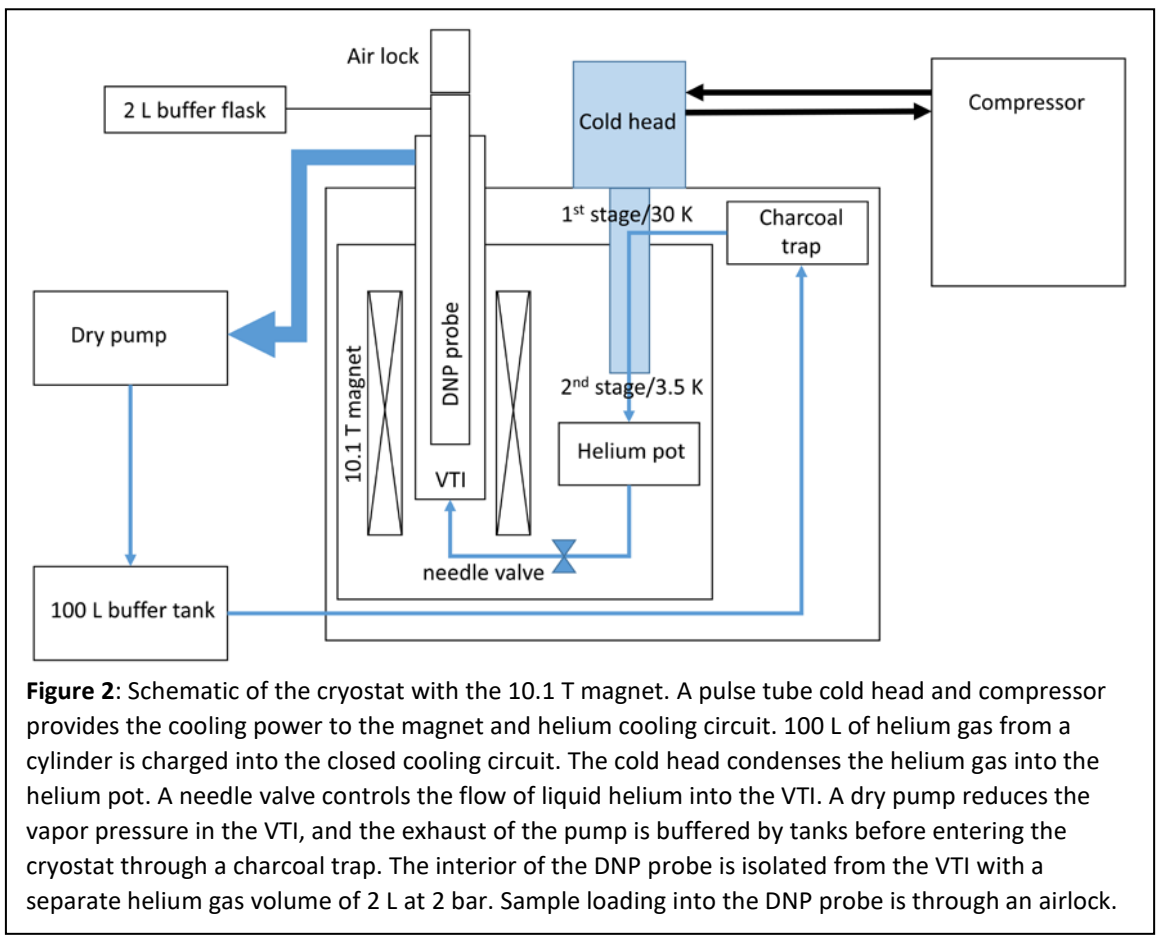

**Figure 2**: Schematic of the cryostat with the 10.1 T magnet. A pulse tube cold head and compressor provides the cooling power to the magnet and helium cooling circuit. 100 L of helium gas from a cylinder is charged into the closed cooling circuit. The cold head condenses the helium gas into the helium pot. A needle valve controls the flow of liquid helium into the VTI. A dry pump reduces the vapor pressure in the VTI, and the exhaust of the pump is buffered by tanks before entering the cryostat through a charcoal trap. The interior of the DNP probe is isolated from the VTI with a separate helium gas volume of 2 L at 2 bar. Sample loading into the DNP probe is through an airlock.

*Magnet*

A 10.1 T cryogen-free superconducting magnet (Cryogenic Ltd, London, UK) provides the static magnetic field, Fig. 1J. The magnet was specified to have a drift rate of <0.1 ppm/h and a homogeneous volume of +/- 25 ppm in a 25 mm tall cylinder of 25 mm diameter. It has fixed current leads and is permanently connected to a SMS120C power supply (Fig 1A: Cryogenic Ltd, London, UK). It is equipped with a persistent mode switch for routine operation. In case of power failure to the compressor (the compressor starts automatically when power returns), the magnet can remain at field for several minutes. Beyond this, the magnet will harmlessly quench. In case of a quench from full field, the magnet is cold within two hours. The magnet is unshielded and provides a strong background field for the hyperpolarized solution. The dissolution is performed 100 mm from the magnet center, where the field is still 80% of the center field.

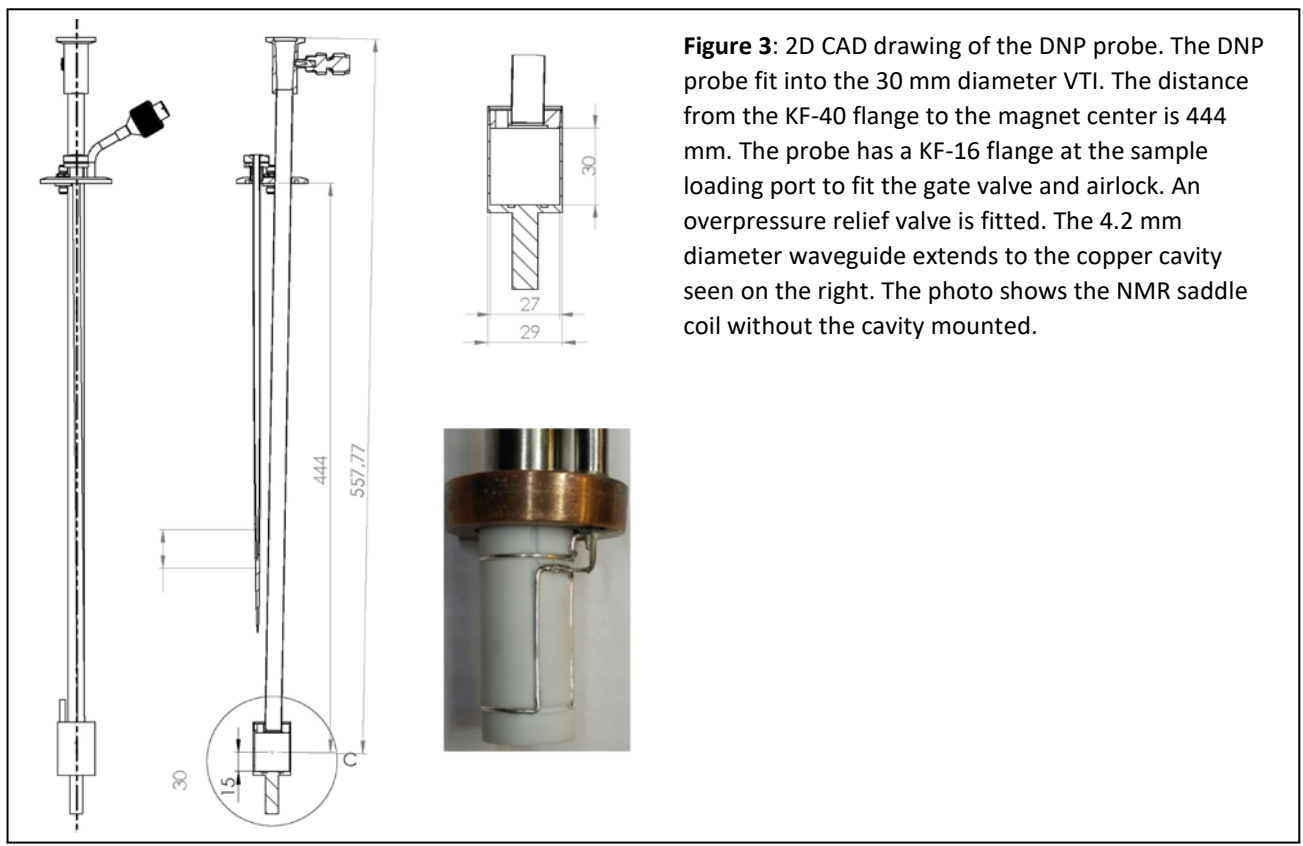

**Figure 3**: 2D CAD drawing of the DNP probe. The DNP probe fit into the 30 mm diameter VTI. The distance from the KF-40 flange to the magnet center is 444 mm. The probe has a KF-16 flange at the sample loading port to fit the gate valve and airlock. An overpressure relief valve is fitted. The 4.2 mm diameter waveguide extends to the copper cavity seen on the right. The photo shows the NMR saddle coil without the cavity mounted.

*DNP probe*

Fig 3 shows a drawing of the DNP probe [27,28]. We designed it for Cryogenic Ltd as a template for similar dDNP systems based on their cryostat and VTI. The DNP probe is isolated from the VTI to avoid contamination of the VTI circuit with air from sample loading and unloading. It consists of a stainless steel upper section mated to a copper MW cavity. The internal of the DNP probe is connected to a buffer volume of 2 L charged with 2 bar of He gas, i.e. 4 L of standard temperature and pressure helium gas. This corresponds to approx. 5 mL of liquid helium, and fills the cavity to half when a sample is loaded. The DNP probe has a Cernox temperature sensor (CX1030, Lakeshore, USA) and a 1 W heater mounted on the cavity. Below 4.0 K the temperature was derived from the VTI helium vapor pressure per ITS-90. The vapor pressure based temperature reading was verified against the Cernox calibration at zero magnetic field. Above 4 K, the Cernox reading was used, ignoring any magnetic field corrections. The MW cavity has an inner diameter of 27 mm and an outer diameter of 29 mm. It has an interior height of 30 mm and is fitted with a saddle coil with 13 mm diameter and 22 mm height [27]. The DNP probe has an approx. 400 mm long, 4.2 mm diameter circular waveguide terminating at the cavity. A UT85SS-CuBe coaxial cable connects the coil to a tune-and-match box external to the VTI. The saddle coil can be single tuned for any frequency up to 428 MHz for $^1$H at 10.05 T, or it can be double tuned for $^{13}$C and $^1$H for the same frequency range.

*MW source(s)*

MW are generated by a 94 GHz source (Fig 1I, Quinstar, USA) with 500 MHz analog tuning range and 300 mW output power with analog attenuation. The source can be connected to either a frequency doubler (D200, Virginia Diodes Incorporated, USA) or tripler (D282, Virginia Diodes Incorporated, USA). The doubler or tripler is biased by -36 V from the power supply (Fig 1E). The efficiencies of the doubler and tripler are 25% and 10%, respectively. Depending on the frequency, the source is connected to a transition from rectangular waveguide (WR10, WR5.1 or WR3.4, respectively) to the circular waveguide (316L stainless steel) of the DNP probe. The transition is vacuum tight with an O-ring against the circular waveguide and with Kapton tape (DuPont, USA) on the input. When the magnetic field is changed, the frequency doubler or tripler and their respective transition is either inserted or removed as required.

*Dissolution system*

The DNP probe sample port has a gate valve (VAT 01224-KA24, Switzerland) mounted between air lock and insertion modules (Fig 1H: GE Healthcare, USA) adapted from the SPINlab polarizer [16]. The airlock is connected to helium purge gas (approx. 1.3 bar) or vacuum (approx. 7 mbar) through the automation valves. A small diaphragm pump (Fig 1Q: KNF, Germany) provides the vacuum. A temperature controller (Fig 1C: Watlow, USA) controls a heater-pressure module (Fig 1G: GE Healthcare, USA), also adapted from the SPINlab, and mounted in proximity to the insertion module. The complete assembly accommodates the use of fluid paths [29–31] available for the SPINlab polarizer, which is the key feature to ensure sterility for human use (the SPINlab also includes a quality control system to enable human studies).

*Automation*

The system is controlled by a real time controller (Fig 1N: cRIO-9035, National Instruments, Austin, TX, USA) running LabVIEW software. The controller has modules for motor control, digital and analogue input and output, and serial communication. A 24 V power supply (Fig 1E) powers the controller. The system runs under computer control with a graphical user interface to control the basic user functions (sample loading, polarization and dissolution). The software logs key parameters such as temperatures, pressures, and sample position from the temperature controller and monitor, as well as pressure gauges. Pneumatics are controlled by a valve block (Fig 1O: Festo, Germany) that delivers compressed air to various pneumatic valves and actuators (Fig 1F). For compatibility with standard house compressed air supplies, the 16 bar pressure required for driving the heater-pressure module is generated by a four times pressure booster (SMC, Japan). High pressure gas, as well as vacuum are controlled with pneumatically actuated high purity diaphragm valves (Fig 1P: Swagelok, Solon, OH, USA).

*NMR*

Solid-state NMR data were acquired with a 400 MHz Unity INOVA NMR console (Agilent, Palo Alto, CA, USA) running VnmrJ 4.2 software. The flip angle was calibrated from a series (hundreds) of NMR spectra acquired with short repetition time (typically 0.1 s). An exponential fit to the signal decay provided the actual flip angle. Solid-state polarization was measured by normalization of the DNP enhanced NMR signal (integral) with the NMR signal at thermal equilibrium at the same conditions. No $^{13}$C background signal from the probe was observed.

Liquid-state polarization measurements were acquired with 400 MHz Direct Drive console running VnmrJ 4.2 software and a 5 mm 1H/X broadband probe (Agilent, Palo Alto, CA, USA). The $^{13}$C Larmor frequency at the three field strengths are 35.9, 71.8 and 108 MHz, respectively. A flip angle of 5° every 5 s was used to measure the decay of the DNP enhanced NMR signal. $T_1$ and initial polarization (at time of dissolution) was calculated from the exponential decay. Liquid state polarization was measured by normalization of the DNP enhanced signal with the NMR signal at thermal equilibrium for the same sample using a 90° flip angle.

*Sample preparation*

[1-$^{13}$C]pyruvic acid (Sigma-Aldrich, Denmark) with the trityl radical AH111501 (GE Healthcare, Denmark) was used in all experiments. The concentration of radical was optimized in coarse increments of 15, 30 and 45 mM at the three magnetic field strengths. 1 mM Gd (Dotarem, Guerbet) doping was tested at the three field strengths for the 15 mM AH11501 sample. Pyruvic acid was loaded into the SPINlab fluid path (GE Healthcare, Denmark) as described in [16,29].

**Results**

The magnet operated in persistent mode at 10.1 T (full field) with a drift rate of -0.06 ppm/h corresponding to 2.8 MHz/week on the MW frequency at 282 GHz. The supplier verified the magnetic field homogeneity at factory. The system cooled from room temperature to operating temperature in less than 24 h and recovered from a quench in less than 2 h. The magnet was energized to full field in less than 30 min.

In persistent mode the base temperature of the VTI was <1.3 K with the pumping capacity of the ACP40 pump and all heat loads (under polarizing conditions). For most of the results presented here, the VTI was maintained at 1.4 K (2.8 mbar helium vapor pressure) by the Lakeshore 350 controller and heater on the probe. Maintaining the probe at fixed temperature permits a steady-state to be established where no adjustments to the helium flow or vacuum pump speed are required. In this mode, reductions in heater power offset MW induced heating and permit stable operation for extended periods. This is in contrast to conventional VTI based systems, which can be operated under a variety of heat loads with only minimal changes in performance. Coupling between the recirculating system and magnet leads to dramatic changes in performance with varying heat loads.

An additional benefit of this operating scheme is the ability to measure the MW power deposited in the sample. The MW power transmission to the cavity at 188 GHz was determined by measuring the heater power required to maintain a fixed temperature as a function of applied MW power, Fig. 4. The power dissipated in the cavity was 5.7 dB (10 log$_{10}$(0.27)) less than the source output power. The loss is consistent with the attenuation that we have measured for circular stainless steel waveguide using a power meter (Ericsson PM-5, Virginia Diodes Inc, USA) at this frequency. This loss increases to 8.5 dB (10 log$_{10}$(0.14)) at 282 GHz.

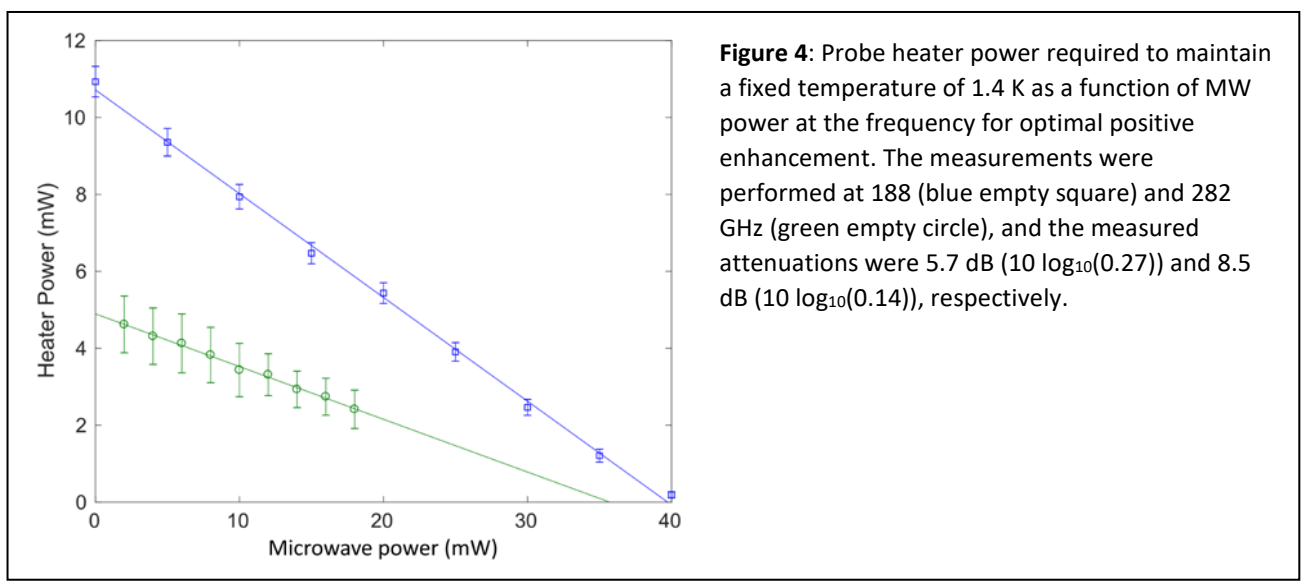

**Figure 4**: Probe heater power required to maintain a fixed temperature of 1.4 K as a function of MW power at the frequency for optimal positive enhancement. The measurements were performed at 188 (blue empty square) and 282 GHz (green empty circle), and the measured attenuations were 5.7 dB (10 log$_{10}$(0.27)) and 8.5 dB (10 log$_{10}$(0.14)), respectively.

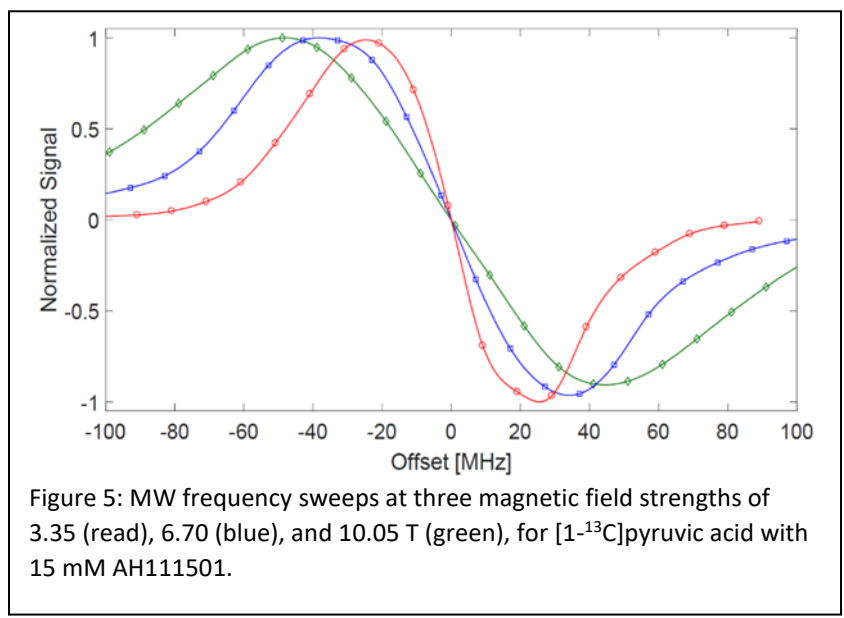

Figure 5: MW frequency sweeps at three magnetic field strengths of 3.35 (read), 6.70 (blue), and 10.05 T (green), for [1-$^{13}$C]pyruvic acid with 15 mM AH111501.

MW frequency sweeps were performed at three magnetic field strengths, 3.35, 6.70, and 10.05 T, Fig. 5. The sweeps were obtained by polarizing for 60 s at each frequency before measuring the NMR signal. The NMR signal was saturated before each polarization period. The separations of the positive and negative extrema are 50, 72, and 93 MHz, respectively. The short polarization period may bias the frequency sweeps, since the build-up time constant is not constant across the DNP spectrum. Therefore, we generally confirmed the sweeps by performing full (several time constants) build-up curves at each MW frequency.

The time constant and final value for DNP build-ups at 282 GHz for [1-$^{13}$C]pyruvic acid with 45 mM AH111501 as a function of MW power is shown in Fig. 6. The figure shows that adequate MW power is available, and that a time constant of about 1800 s (30 min) can be obtained for [1-$^{13}$C]pyruvic acid. In contrast to the MW frequency sweeps, the MW power sweeps are very biased towards higher power, if short build-up periods are used. The data reported here is obtained by fitting of full build-up curves at each MW power.

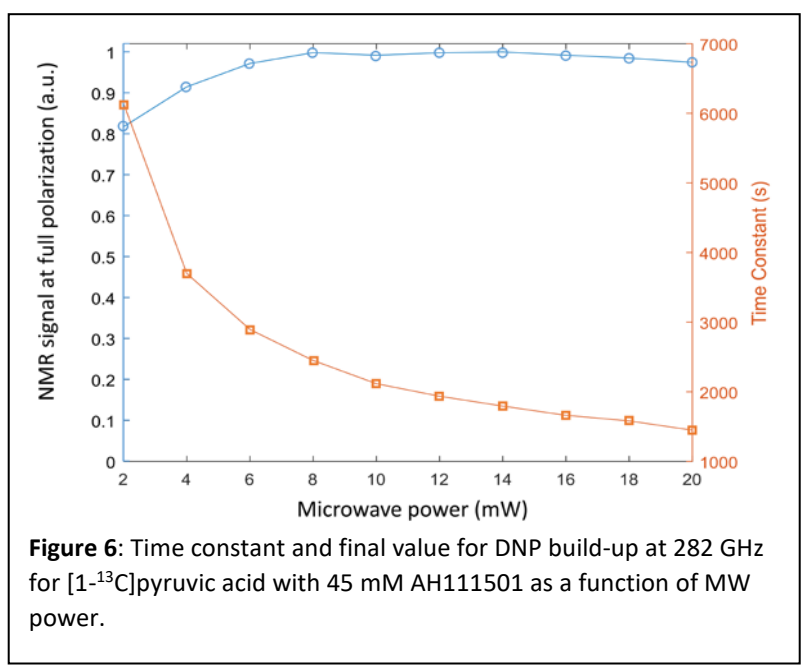

Figure 6: Time constant and final value for DNP build-up at 282 GHz for [1-$^{13}$C]pyruvic acid with 45 mM AH111501 as a function of MW power.

Examples of DNP build-up curves at 94, 188, and 282 GHz for [1-$^{13}$C]pyruvic acid for close to optimal conditions (MW frequency, modulation and power) are shown in Fig. 7. The radical concentration has been varied to approach optimal conditions, and a higher radical concentration is required at higher magnetic field to obtain maximum polarization with the shortest possible build-up time constant. The 15 mM AH111501 sample has a $^{13}$C nuclear relaxation time, $T_1$, of 90,000 (25 h), 33,000 (9.2 h) or 6,400 s at 10.1, 6.7 and 3.35 T, respectively. The DNP build-up time constant at the three field strengths are 26,000 (7.2 h), 3,700 and 1,000 s, respectively, for the 15 mM AH111501 concentration. A build-up time constant of approx. 1200 s was obtained at 6.7 T for 30 mM AH111501. The build-up curves have been obtained by acquiring a spectrum with a 1° flip angle every 300 s. At the two highest field strengths, MW modulation was applied at a rate of 1 kHz and amplitude of 25 or 50 MHz. No effect of modulation was seen at 3.35 T, whereas the effect was pronounced at 10.1 T with a two to three times increase in the DNP enhancement for 20 to 50

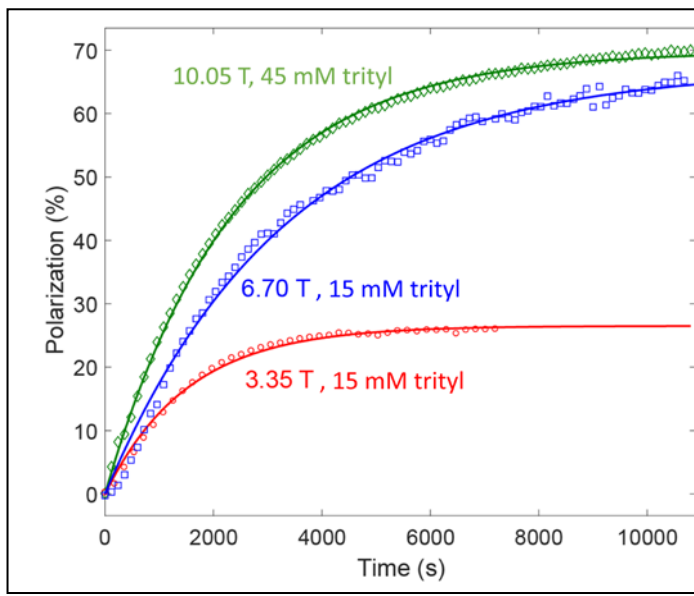

Figure 7: DNP build-up curves at 94, 188, and 282 GHz for [1-$^{13}$C]pyruvic acid for close to optimal conditions (MW frequency, modulation and power). The radical concentration has been varied to approach optimal conditions.

MHz modulation amplitude. At the higher radical concentrations, the effect of MW modulation was significantly reduced, but still provided an approx. 50% improvement at 10.1 T. MW modulation had a similar effect on the build-up time constant as on the final polarization. We estimate an uncertainty of ±5% based on the three replicates for all numbers. The fitting uncertainties were typically 1% or less.

The samples in Fig. 7 have been dissolved with liquid-state polarization of 27% at 3.35 T, 70% at 6.7 T, and 70% at 10.1 T. The standard deviation on the liquid-state polarization is ±5% with three replicates at each magnetic field strength.

Fig. 8A shows a representative sample loading profile (position within the DNP probe; -630 mm is the magnet center) and corresponding temperature of the DNP probe for a 1.5 g sample. With such a large sample, heat loads are at the upper end of the operating range and the trends most pronounced. The loading profile has been empirically optimized to impose a low, but consistent heat load on the helium bath throughout the loading process. The temperature at the first stop at -400 mm is less than 70 K, ensuring that the sample freezes rapidly. If the sample has been frozen prior to the loading process, this pause could be shortened. Smaller samples can be loaded considerably faster, with the loading profile for a 50 mg sample taking less than 3 min. Fig 8B shows the DNP probe temperature during a dissolution of the 1.5 g sample. The dissolution takes place at -530 mm, 100 mm above the magnet center, and is retracted immediately at the end of the dissolution. A peak temperature of the DNP probe is 1.9 K, which settles to base temperature within two minutes. A smaller sample would have less impact on the DNP probe temperature. Thus, sample loading and unloading (dissolution) adds between 5 and 15 min to the polarization cycle depending on sample size, thus, exceeding SPINlab and adding fractional to the polarization cycle.

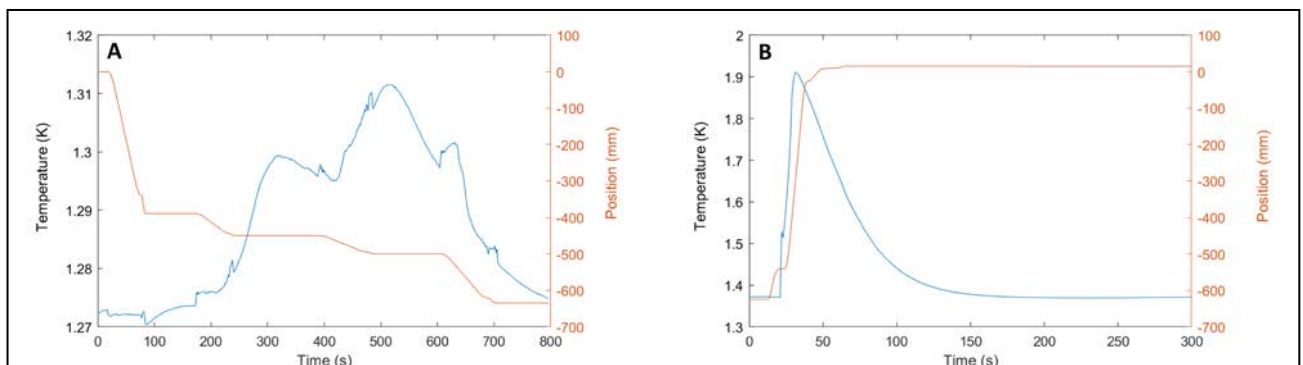

Figure 8: A. Temperature profile of the helium bath (DNP probe) during loading of 1.5 g of [1-$^{13}$C]pyruvic acid with 15 mM AH111501. The sample is lowered gradually according to an insertion profile that has been empirically adjusted to minimize the heat load to the helium bath. Sample position is indicated on the secondary ordinate. B. Temperature profile of the helium bath (DNP probe) during a dissolution of the same sample. The sample is raised 100 mm before dissolution, and immediately retracted when the dissolution is complete.

**Discussion and conclusion**

In comparison to NMR magnet design, there are some notable differences in requirements for dDNP systems. One such area is the desirable effect of stray magnetic field. The dissolution process benefit from taking place at high magnetic field in principle. Furthermore, the dissolved sample must be handled in a background magnetic field to avoid low field relaxation and non-adiabatic field changes. The most straightforward manner to accomplish these goals is to eschew the use of active shielding, which may produce abrupt magnetic field gradients, and thereby maximize the stray field. Another consideration is minimization of the length of the VTI. On a traditional cryogen cooled magnet, this distance is largely defined by the cryogen vessels contained within the cryostat that require a certain volume and benefit from long heat conduction paths. As a cryogen-free magnet is free of these constraints, it is possible to reduce the height of the VTI dramatically. Therefore, the hyperpolarized solution is always within 1 m of the magnet center and the background field never drops below 2 mT (when at 6.7 T or above).

A further advantage of this is a concomitant reduction in the overall height of the system, a lower weight and cost, and the ability to take the magnet off field for periods, and within hours or days bring the polarizer operational. One consequence of shortening the VTI is a relatively high thermal gradient along the length of the VTI. This thermal gradient must be considered when designing the probe as it is possible to inadvertently introduce large heat loads through e.g. conduction or thermoacoustic oscillations.

With regard to the magnet itself, requirements for dDNP are generally less demanding than those for NMR. In particular, magnetic field homogeneity and drift requirements are relaxed by several orders of magnitude as compared to high-resolution NMR magnets. One caveat is that the homogeneity requirements must be met without the use of passive shims, which are optimized for use at a single magnetic field strength. Independent superconducting shims are best avoided, as they would need to be ramped at the same time as the main magnetic field. While drift requirements are relatively relaxed (0.1 ppm/hr), the magnet must be run in persistent mode, as the drift rate of commercially available magnet power supplies is not able to meet this specification. Furthermore, persistent mode reduces the thermal load on the magnet by eliminating resistive heating in the current leads. The reduced thermal load increases the power available for helium liquefaction.

The obtained drift rate of the magnet is acceptable for weekly calibration of the optimal MW frequency even for radicals with narrow ESR linewidth. With a tuning range of approx. 2 GHz (at 10.1 T) for the MW source, magnetic field correction would be required less than annually. We have implemented an automated procedure for resetting the magnetic field and typically perform this monthly. The large homogeneous volume was specified to allow multiple human doses to be polarized simultaneously. In this work, we only polarized a single sample of full human dose equivalent (1.5 g of pyruvic acid), but it should be possible to fit three human dose samples (SPINlab fluid paths) into the DNP probe and into the homogeneous volume of the magnet.

Another key thermal consideration is the heat introduced in the sample loading and dissolution processes. These processes are critical, because they have the ability to introduce a sudden heat load that will rapidly propagate through the system, increase the temperature of the second stage of the cold head, and potentially quench the magnet. We have demonstrated, that even under the most demanding conditions, will the sample loading and dissolution process introduce little heat into the system. The sample vial is lifted 100 mm prior to dissolution and immediately removed after the dissolution process is complete. However, the loading process must be conducted with care as it has the potential to evaporate significant quantities of liquid helium. It is preferable to direct as much of the heat from the sample to the enthalpy of the helium gas instead of the helium bath. This is more easily accomplished in this system due to the higher operating temperature in comparison to the SPINlab polarizer. The higher operating temperature leads to a higher operating pressure, leading to more efficient thermal transfer between the sample and the cold helium gas, i.e. faster cooling rate of the sample. In the case of the SPINlab, the challenge is to reject the heat to the cold head to preserve the finite helium volume. For this system, the limitation is the capacity of the cold head to condense the warm exhaust helium. The most popular substrate for hyperpolarized metabolic MR, [1-$^{13}$C]pyruvic acid, was studied in the work. We show that the previously reported strong field dependence in the range 3.35 to 4.6 T [32,33] does not extrapolate to higher magnetic fields; up to 10.1 T. It seems that, for this sample, the maximum achievable polarization is approx. 70% in agreement with other work at high magnetic field strength [11].

The g-tensor for the OX063 trityl (AH111501 is the methylated OX063) was determined by Lumata et al [34] to be axially symmetric with $g_\perp$ = 2.00319(3) and $g_\parallel$ = 2.00258(3. Thus, the $g_\parallel$- $g_\perp$ is -0.00061, which corresponds to a spectral width of 8.5 MHz/T. This is less than the $^{13}$C Larmor frequency of 10.71 MHz/T. Electron-electron dipolar broadening contributes to the line width, in a first approximation linearly with concentration. Chen et al [35] measured the trityl line width in frozen solution for 20, 40 and 60 mM at X-band (minimal contribution from g-anisotropy). They found that the line shape changed towards Lorentzian at the highest concentration, and that the line width was less than the dipolar coupling of 20 MHz between a minimal distance (approx. 1 nm) trityl spin pair. The 8.5 MHz/T g-anisotropy, convoluted with a dipolar broadening of 5-10 MHz at 45 mM corresponds well with the peak-peak separation of the DNP spectrum of 93 MHz, which is significantly less than the $^{13}$C Larmor frequency of 107.1 MHz at 10.1 T. We therefore conclude that the ESR spectral width of the trityl is well matched for direct $^{13}$C DNP via a CE or TM mechanism up to approx. 6.7 T, but above this field strength, DNP via these two mechanisms become less efficient due to a too narrow ESR line relative to the $^{13}$C Larmor frequency. Consequently, we observe that a much higher radical concentration is required when the magnetic field strength increases from 3.35 to 10.1 T in order to broaden the ESR spectrum. Thus, the optimal radical concentrations are 15, 30 and 45 mM at 3.35, 6.7 and 10.1 T, respectively. Furthermore, the higher radical concentration makes the ESR line more homogeneous. At the lowest radical concentration, 15 mM, we observed a significant increase of the DNP enhancement and shortening of build-up time constant with modulation of the MW frequency. The effect of MW frequency modulation was significantly offset when the radical concentration was increased. To the point that MW frequency modulation had little effect at 6.7 T at 30 mM, but still significant effect at 10.1 T and 45 mM. Thus, at high field the ESR line becomes inhomogeneous as expected, and the DNP mechanism shifts from TM to CE. Increased radical concentration is unable to make the ESR line effectively homogeneous.

Both Lumata et al and Chen et al studied the electron longitudinal relaxation time, $T_{1e}$, for trityls as a function of temperature and magnetic field. A temperature dependence of approx. linear was reported, characteristic of a direct process, which favors increasing the temperature to shorten $T_{1e}$. Surprisingly, they reported almost no magnetic field dependence as would be expected for the dominating Orbach and direct mechanism, implying that $T_{1e}$ should not be a bottleneck at higher magnetic field strengths either. This is supported by the fact that we have not observed any benefit of Gd doping at the two highest magnetic field strengths for this sample.

Since the nuclear relaxation time, $T_1$, continues to increase with approx. the square of the magnetic field strength up to 10 T, it seems that the leakage factor cannot explain the stagnation at 70% polarization. The nuclear longitudinal relaxation time, $T_{1n}$, is due to the presence of trityl. The direct dipolar relaxation by the electron spin is modulated by the electron spin lattice relaxation time ($T_{2e}$ on the time scale of tens of nanoseconds) [9]. Since we would expect $\omega\tau \gg 1$ at low temperature and high field, the relaxation rate should follow a $B_0^{-2}$ dependence. This is consistent with our observation of $T_{1n}$ increasing approx. quadratic with magnetic field. However, this is probably coincidental, since this relaxation mechanism, electron-nucleus dipolar relaxation, should further scale with the $(1-P_{0e}^2)$ factor. At 1.4 K and 3.35, 6.7 and 10.1 T, $P_{0e}$ is 92.3%, 99.7% and 99.99%, respectively, and, thus, the factor becomes 0.147, 0.00632, 0.000253, respectively. This is approx. a factor 23 from 3.35 to 6.7 T, and another factor 25 from 6.7 to 10.1 T. A more reasonable interpretation of the radical induced nuclear $T_1$, is that it proceeds through the same three spin transitions that give rise to the DNP enhancement, indirect nuclear relaxation [9]. Similar to the discussion above for DNP, the observation that the $^{13}$C Larmor frequency starts to exceed the ESR line width, makes these transitions less probable. Furthermore, the non-linear shortening of $T_{1n}$ with radical concentration as it increases from 15 to 45 mM supports this interpretation. In [36] we observed an almost linear shortening of $T_{1n}$ with radical concentration at 3.35 T and 1.2 K, and no effect of Gd doping on $T_{1n}$. However, another effect may be that small amounts of clusters of trityls form fast relaxing centers [35] that shorten $T_{1e}$. Their implication for $T_{1n}$ and DNP remains to be further investigated.

In general, a technical limitation for high field DNP, has been the availability of MW sources. However, improved waveguide transmission, the short DNP probe, and efficient multipliers has extended the accessible range. In this work, we show that this is no longer a limitation of high field dDNP. At 10.1 T approx. 10 mW is required for maximal DNP and shortest build-up time constant. At 6.7 T the required MW power is slightly higher, approx. 20 mW, to achieve fastest build-up, and at 3.35 T the optimal MW power had increased to approx. 40 mW. These numbers are influenced by technical performance, but illustrates the general trend that the narrower ESR line at the lower magnetic field strengths absorb more MW power and lead to faster DNP time constants. In principle, any field strength in the range defined by the cut-off frequency of the circular waveguide (approx. 42 GHz) and the maximum field strength of 10.1 T can be used. In recent years, MW sources with good properties for DNP at 196 GHz (7 T) and 263 GHz (9.4 T) have become available.

The HYPERMAG polarizer provides several important features. It runs in continuous mode (in contrast to the SPINlab that requires overnight regeneration of the contained helium) without consumption of any cryogens. It offers the widest range of operating conditions for dDNP, up to 10.1 T magnetic field strength and any temperature from room temperature down to 1.3 K. The system is very convenient, with a high level of automation for the user, and reliable. A liquid state polarization of 70% was obtained for [1-$^{13}$C]pyruvate with a solid state build-up time constant of approx.1200 s (20 min) allowing a throughput of at least one sample per hour including sample loading and dissolution We have built two systems, the first has been running for 15,625 h/651 days (compressor running hours) and the second has been in operation at Dr Mikko Kettunen, University of Eastern Finland since Aug, 2017. In conclusion, using a custom dry magnet, cold head and recondensing, closed-cycle cooling system, combined with a modular DNP probe, automation and fluid handling systems; we have designed a unique dDNP system with unrivalled flexibility and performance.